\documentclass[aps,showpacs,amsmath,amssymb,superscriptaddress,pre,preprint]{revtex4-1}
\usepackage{graphicx}
\usepackage{amsmath}
\usepackage{hyperref}
\usepackage{dcolumn}
\usepackage{bm}
\usepackage{color}
\usepackage{nicefrac}

\begin{document}

\title{Spontaneous symmetry breaking in finite systems and anomalous order-parameter correlations}

\author{F.~K.~Diakonos}
 \email[]{fdiakono@phys.uoa.gr}
\affiliation{Faculty of Physics, University of Athens, GR-15784 Athens, Greece}

\author{Y.~F. Contoyiannis}
 \email[]{yiaconto@uniwa.gr}
\affiliation{Department of Electrical and Electronics Engineering, University of West Attica, GR-12244, Egaleo, Greece}
\affiliation{Faculty of Physics, University of Athens, GR-15784 Athens, Greece}

\author{S.~M. Potirakis}
\email[]{spoti@uniwa.gr}
\affiliation{Department of Electrical and Electronics Engineering, University of West Attica, GR-12244, Egaleo, Greece}

\date{\today}

\begin{abstract}

We show that the spontaneous symmetry breaking can be defined also for finite systems based on the properly defined jump probability between the ground states in the 2d and 3d Ising models on a square and a cubic lattice respectively. Our analysis reveals the existence of an interval in the temperature (control parameter) space within which the spontaneous symmetry breaking takes place. The upper limit of this region is the pseudocritical point where the symmetric vacuum bifurcates in energetically degenerate non-symmetric vacua, initiating the spontaneous symmetry breaking process. The lower limit, identified as the temperature value at which the spontaneous symmetry breaking is completed, is characterized by maximal characteristic time for the decay of magnetization (order parameter) auto-correlations. We argue that this anomalous enhancement of auto-correlations is attributed to the transition from type I to on-off intermittency in the order parameter dynamics. Possible phenomenological implications of this behaviour are briefly discussed. 

\end{abstract}

\pacs{} 

\maketitle
Critical phenomena are the origin of the physical world's structure at all scales. Their realization in physical systems in usually related to the spontaneous breaking of symmetries implied by the parametric dependence of the vacuum state of a system on  control parameters. According to this scenario a symmetry respecting ground state of a physical system, becoming dynamically unstable, bifurcates in two (or more) dynamically stable ground states whenever a critical value of the control parameter (critical point) is crossed. The emerging ground states form a degenerate space and the original symmetry transform, leaving the dynamics of the system invariant, generates a mapping between the different ground states. Then, the underlying symmetry is spontaneously broken since the ground states do not remain invariant under the corresponding transformation. Although this scenario is rigorously defined in infinite systems, it does not apply to real physical systems which posses finite size. In fact, in a finite system a symmetry can never break spontaneously \cite{Binder1985} in the strict mathematical sense. Obviously, the spontaneous breaking of a symmetry in a finite system cannot take place at the pseudocritical point signalling the occurrence of the bifurcation of the vacuum state, which is a necessary but not sufficient condition. In fact, well after the crossing of the bifurcation point in the control parameter space, the system behaves as having a ground state which is still symmetric, since the new vacua, which emerge, communicate through the underlying dynamics. In this sense the physical system is in a symmetric superposition of the corresponding vacua, a term borrowed by quantum mechanics. Effectively, this quantum mechanical picture is always valid also for a classical field (like the order parameter density) defined in a finite spatial region. This is due to the fact that there is a finite jump probability from the one ground state to the other when the system possesses a fixed size.

The present work aims to present a definition of spontaneous symmetry breaking (SSB) in a finite system. To achieve this task, we study the jump probabilities between the new vacua occurring after crossing the pseudocritical control parameter value. We show that these probabilities, as a function of the control parameter, posses an exponential suppression term. The latter defines a characteristic control parameter value, after which, the jump probability becomes negligibly small and therefore the symmetry can be considered as broken. In practice, this situation is realized in nature since the time interval during which an observer probes a finite system is also finite. We will call this characteristic control parameter value as the {\it SSB completion point}. Thus, the pseudocritical point and the SSB completion point define an interval in the control parameter space within which the symmetry, despite the emergence of the new vacua, remains unbroken. Departing from this interval in the control parameter space, the communication between the new vacua, through the order parameter dynamics, is practically lost for finite observation time. The width of the control parameter region, within which the completion of the SSB takes place, depends on the size of the system and increases as the latter decreases. We will in the following refer to this region as the {\it completion region of SSB}. Analyzing the order parameter dynamics in the SSB completion region we reveal an anomalous behaviour of the corresponding auto-correlations expressed by an enhancement in the border of this region. Furthermore, employing the Method of Critical Fluctuations (MCF) \cite{Contoyiannis2002} to analyse the order parameter time-series at different control parameter values within the SSB completion region, we observe a transition of the order parameter dynamics from intermittency type I \cite{Manneville1980, Schuster1995} to the on-off intermittency regime \cite{Platt1993}. Finally, as we discuss at the end of this letter, despite its theoretical character, our proposal should be relevant for the determination of the critical point in systems which are microscopic and possess a finite lifetime. Such a case occurs, for example, in the search of the QCD critical endpoint in the freeze-out state of the fireball formed in relativistic ion collisions \cite{Gavai2016}. In this context our findings can explain eventual discrepancy between experimental and simulated results concerning the critical point location \cite{Gavai2016,Ratti2018}.

To further illustrate our claims it is necessary to employ a concrete system. Here we will consider the magnetization dynamics of the ferromagnetic Ising model with next neighbour interactions in 2 and 3 dimensions, defined in a square and cubic lattice respectively. It is given by the Hamiltonian:

\begin{equation} H=-J \displaystyle{\sum_{\langle i,j \rangle}} s_i s_j~~~~~;~~~~~J > 0
\label{eq:eq1}
\end{equation}
where $\langle i,j \rangle$ denotes summation over next neighbours. As usual, the coupling $J$ can be absorbed into the inverse temperature $\beta=T^{-1}$. The system obeys periodic boundary conditions. The simulations are performed in lattices of relatively small size to reveal the phenomenon addressed previously in the clearest way. The Metropolis algorithm is used in all simulations, taking into account autocorrelations in order to produce statistically independent spin configurations. Our calculations are performed using $\mathcal{N}=2 \cdot 10^6$ sweeps/spin while we have checked the convergence of our results for specific values of the size $L$ and the temperature $T$, in both the 2d as well as the 3d model, using several $\mathcal{N}$ values ranging from $10^5$ to $10^7$ sweeps/spin.

We focus on the properties of the average magnetization $M=\displaystyle{{\sum_{i} s_i \over \Omega}}$ at equilibrium. With $\Omega$ we denote the volume of the considered lattice. Firstly, we investigate the distribution $P(M)$ for various temperature values. In Fig.~1a we show $P(M)$ at $\beta^{-1}=2.3$ for a square lattice of $128 \times 128$ size. This temperature value corresponds to the pseudo-critical one, notated as $T_{pc}$, at which a plateau region around $M=0$ is formed in $P(M)$. This property reflects the fact that the fixed point $M=0$ becomes marginally stable as an intermediate step in its transition from a stable to an unstable one with decreasing temperature. In Fig.~1b we show $P(M)$ for $\beta^{-1} = 2.27$ where the formation of the two new stable minima in the magnetization effective potential $V_{eff}(M)$ is clearly seen. Notice that the minima of $V_{eff}(M)$ appear as maxima in the distribution $P(M)$ since $P(M) \propto e^{-\Omega V_{eff}(M)}$. However, despite the presence of these two minima in $V_{eff}(M)$, the dynamics in the magnetization space are still respecting the $Z(2)$ symmetry, since there is finite, symmetric in the direction, probability to jump from the one minimum to the other. For a clearer display of this behaviour we include in this plot an inset, presenting the magnetization time-series after equilibrium is reached. The jumps from the left stable minimum to the right and vice versa are clearly observable in this inset. 

\begin{figure}[htbp]
\includegraphics[width=0.8\columnwidth]{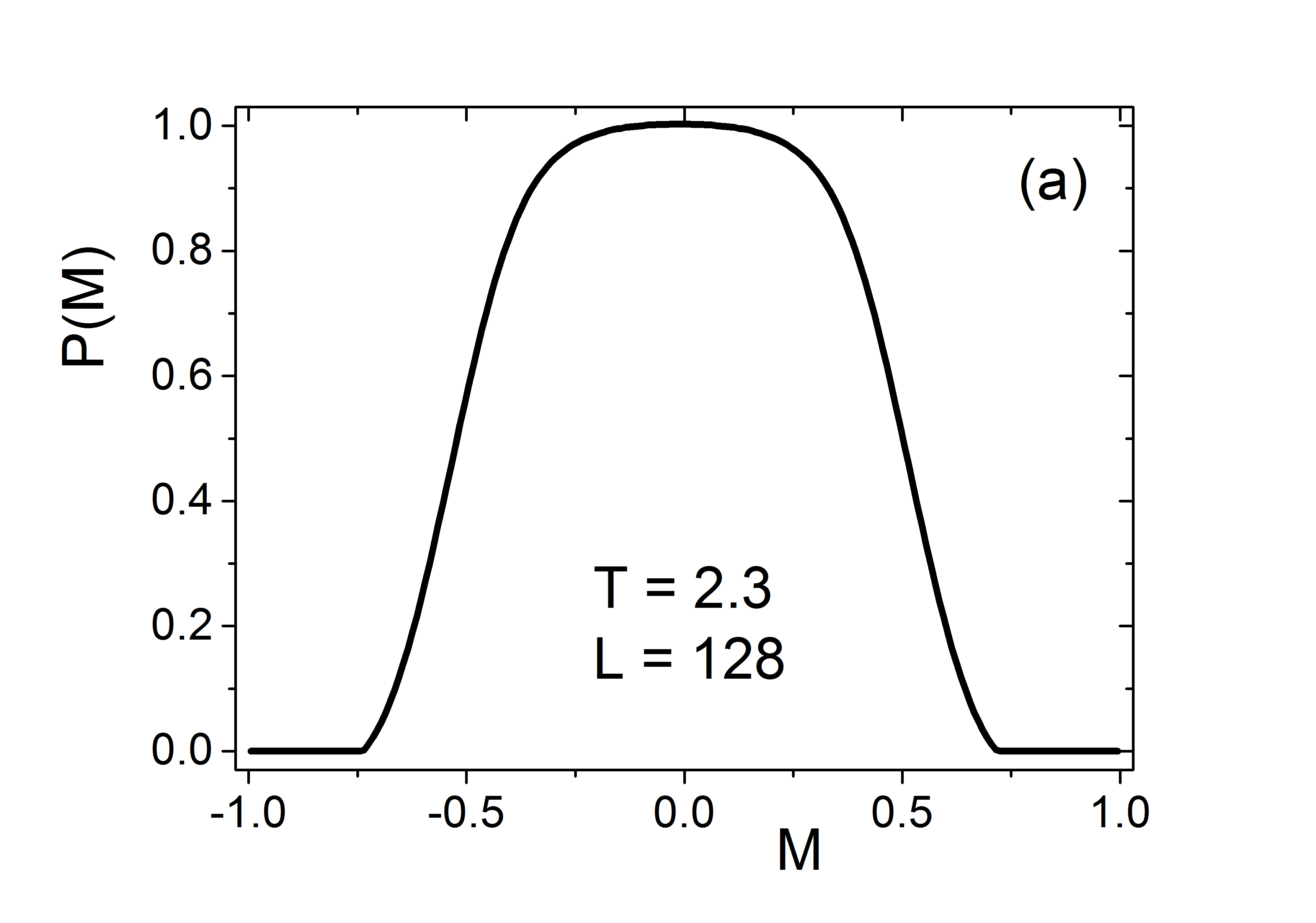}
\includegraphics[width=0.8\columnwidth]{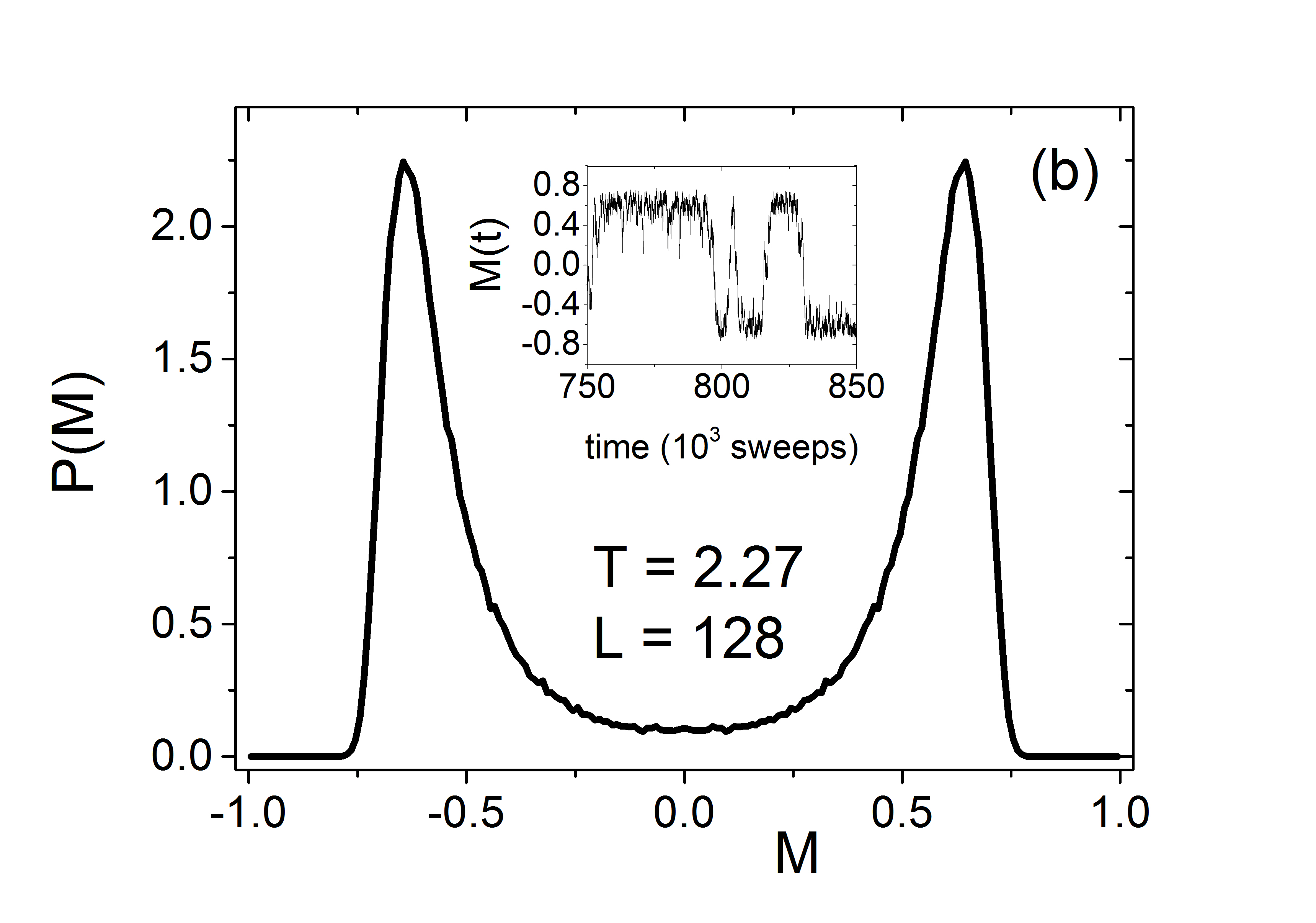}
\caption{The magnetization distribution in 2d Ising square lattice with $L=128$ and periodic boundary conditions for: (a) $T \approx T_{pc}=2.3$ and (b) $T \approx T_{SSB}=2.27$. In the inset in (b) is shown a part of the corresponding magnetization time-series.}
\label{fig:fig1}
\end{figure}

In the following we define as left-right and right-left jump probabilities $p_{LR}$ and $p_{RL}$ as:
\begin{equation}
p_{LR}=\displaystyle{\frac{n_{-+}}{\mathcal{N}}}~~~~~~;~~~~~~p_{RL}=\displaystyle{\frac{n_{+-}}{\mathcal{N}}}
\label{eq:2}
\end{equation}
where $n_{-+}$ ($n_{+-}$) is the number of sign changes of the magnetization $M$ from negative (positive) to positive (negative) values respectively, in a given number of time steps $\mathcal{N}$ measured in sweeps. Since in the considered temperature range, due to $Z(2)$ symmetry, it holds $p_{LR}=p_{RL}$, we restrict our study on the jump probability $p_{RL}$ without loss of generality. Our goal is to calculate $p_{RL}$ at different subcritical temperatures close to the pseudocritical $T_{pc}$ and for different lattice sizes $L$ to obtain $p_{RL}(L,T)$. The results of our calculations are presented in Figs.~2(a,b). In Fig.~2a we show the function $Q(T,L)=-{1 \over b(L)}\ln\displaystyle{\left(\frac{p_{RL}(T,L)}{p_{RL}(T_{pc}(L),L)}\right)}$ for the 2d Ising model and various sizes of the square lattice $L=64$ (black open circles), $L=80$ (red crosses), $L=96$ (green up triangles), $L=112$ (blue stars) and $L=128$ (violet triangles down). The scaling factor $b(L)$ is used to reveal the universal behaviour obeyed by the calculated jump probabilities. On the x-axis we set the temperature difference $T_{pc}(L)-T$ with $T_{pc}(L)$ the pseudocritical temperature corresponding to each $L$-value, calculated with separate simulations and cross-checked with the literature values \cite{Malakis2004,Kaupuzs2011}. We observe that the results for the different lattice sizes are concentrated on a single curve providing a universal law for the decay of the jump probability with temperature in the immediate neighbourhood of the pseudocritical point within the subcritical region. The dashed black line is the function $(T_{pc}-T)^{{3 \over 2}}$. As shown in Fig.~2b, similar results for $Q(T,L)$ are also obtained for the 3d Ising case using cubic lattices with $L=16$ (black open circles), $L=20$ (red crosses), $L=24$ (green up triangles), $L=28$ (blue stars) and $L=32$ (violet triangles down). 

\begin{figure}[htbp]
\includegraphics[width=0.8\columnwidth]{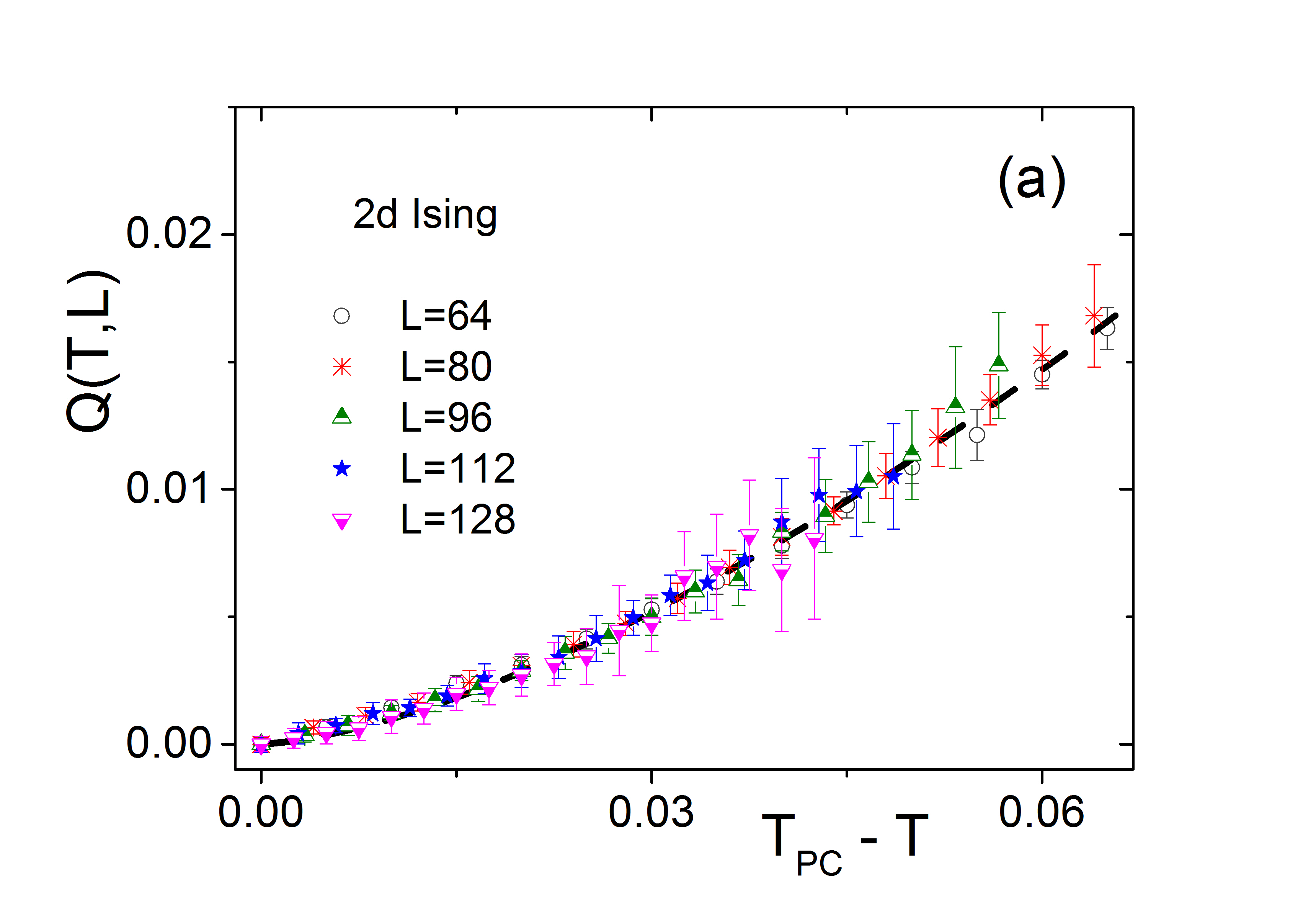}
\includegraphics[width=0.8\columnwidth]{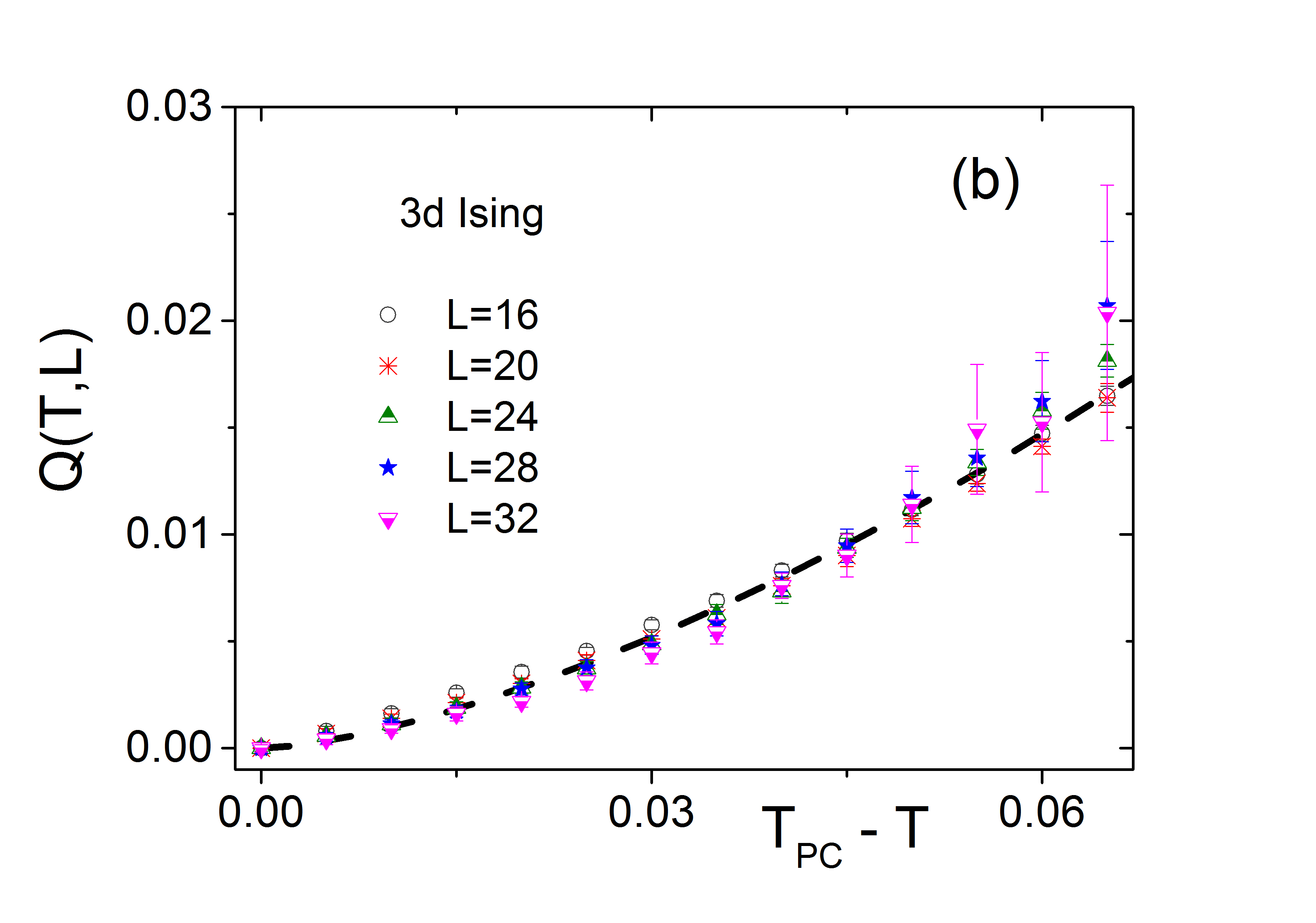}
\caption{The quantity $Q(T,L)=-{1 \over b(L)}\ln\displaystyle{\left(\frac{p_{RL}(T,L)}{p_{RL}(T_{pc}(L),L)}\right)}$ as a function of $T_{pc}-T$ for: (a) the 2d Ising model using lattices with size $L=64,~80,~96,~112,~128$ and (b) the 3d Ising model using lattices with size $L=16,~20,~24,~28,~32$. The scaling factors $b(L)$ in each case are obtained though fitting.}
\label{fig:fig2}
\end{figure}

The universal behaviour in Figs.~2a and 2b leads to a modified exponential decay law given by the function:
\begin{equation}
p_{RL}(T,L)=p_{RL}(T_{pc}(L),L) e^{-b(L) (T_{pc} - T)^{{3 \over 2}}}
\label{eq:3}
\end{equation}
The characteristic exponent $b(L)$ contains a temperature scale which determines the characteristic temperature $T_{SSB}$ below which the jump probability $p_{RL}$ becomes negligibly small in a lattice of size $L$. In fact we can write $b(L)=\frac{\sigma(L)}{T_{pc}^{{3 \over 2}}}$ with $\sigma(L)$ dimensionless. The jump probability becomes negligible small when $\sigma(L) \left(\frac{T_{pc}-T}{T_{pc}}\right)^{{3 \over 2}} \approx \left( \frac{8 \Gamma[{2 \over 3}]}{3}\right)^{3 \over 2}$  ($\Gamma[z]$ is the Gamma function). This condition is obtained through a linear approximation of the distribution $p_{RL}(T,L)$ crossing the function of Eq.~(\ref{eq:3}) at its maximum curvature point. In this limit $T \approx T_{SSB}$ leading to:
\begin{equation} 
T_{SSB}\approx T_{pc} \left(1 - \frac{8 \Gamma[{2 \over 3}]}{3}\left(\frac{1}{\sigma(L)}\right)^{{2 \over 3}} \right)
\label{eq:4}
\end{equation}
It has been checked that $T_{SSB}$ calculated through Eq.~(\ref{eq:4}) fulfils the inequality
$p_{RL}(T_{SSB},L) \mathcal{N} < 0.5$, which in turn means that there is no jump event between the two free-energy minima within the simulation time-interval $\mathcal{N}$. As shown in Fig.~3 the coefficient $\sigma(L)$ has a power-law dependence on $L$ in the form $\sigma(L) \sim L^2$ for the 2d Ising and $\sigma(L) \sim L^{\frac{7}{3}}$ in the 3d Ising case.
\begin{figure}[htbp]
\includegraphics[width=0.95\columnwidth]{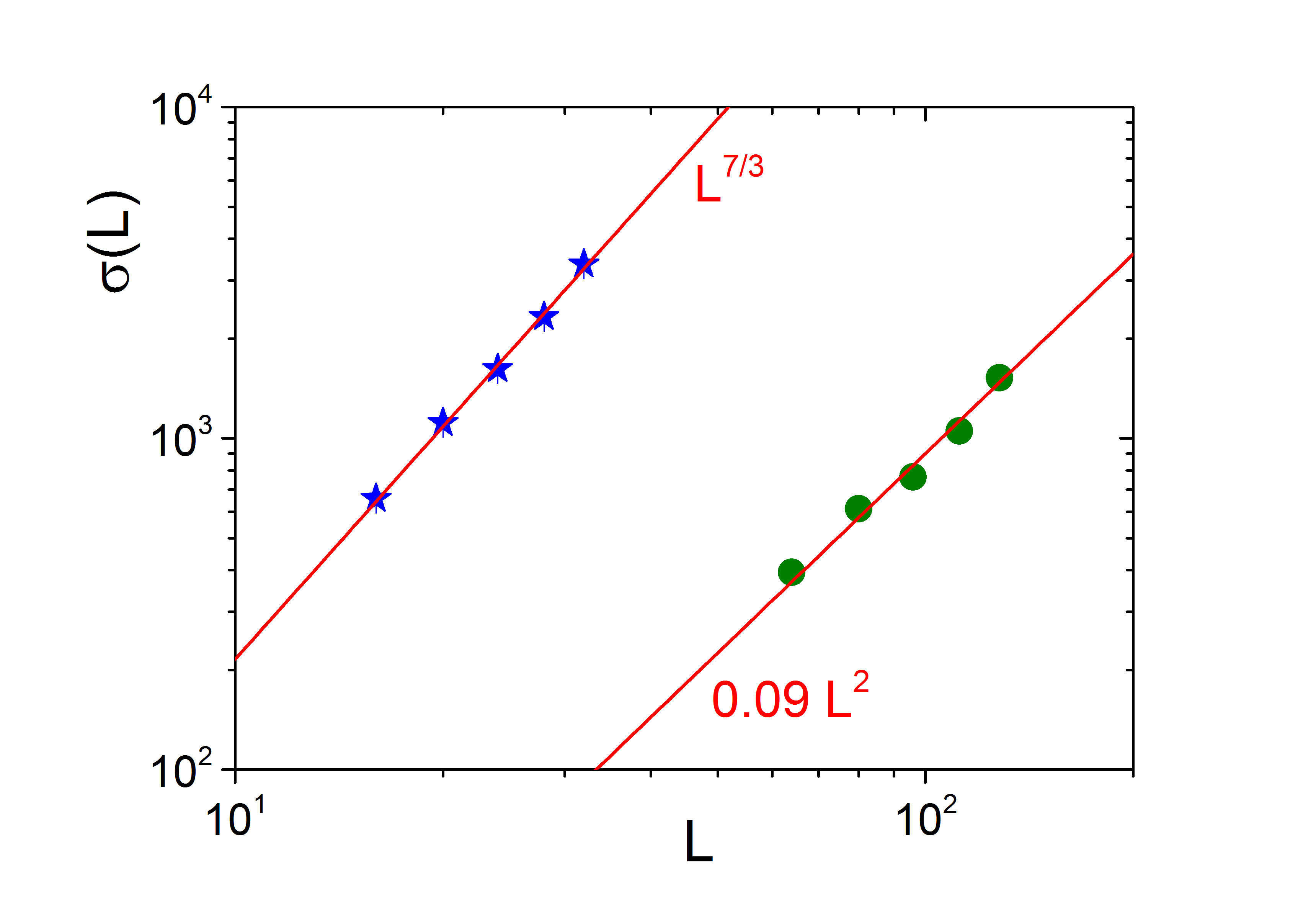}
\caption{The coefficient $\sigma(L)$ used in Eq.~(\ref{eq:3}) ($b(L)=\frac{\sigma(L)}{T_{pc}^{3/2}}$) as a function of $L$ for the 2d Ising (olive circles) and the 3d Ising (blue stars) case. The red lines, representing the scaling laws $0.09 L^2$ (2d Ising) and $L^{{7 \over 3}}$ (3d Ising), are displayed to guide the eye.}
\label{fig:fig3}
\end{figure}
Combining Eq.~(\ref{eq:4}) with the results shown in Fig.~3 we obtain that the size $\Delta T = T_{pc}-T_{SSB}$ of the SSB completion region in temperature space depends on the lattice size $L$, shrinking with increasing $L$ according to the power-laws:
\begin{equation}
\Delta T_{2d} \sim L^{-4/3}~~~~;~~~~\Delta T_{3d} \sim L^{-14/9}
\label{eq:5}
\end{equation} 

Our next goal is to explore in more detail the magnetization dynamics within the SSB completion region. In a previous work we have shown, in the context of the 3d Ising model, that the magnetization dynamics at the pseudocritical point possesses characteristics of type I intermittency, expressed by a power-law form of the distribution $N(\lambda) \sim \lambda^{-\frac{\delta + 1}{\delta}}$ ($\delta$ is the isothermal critical exponent) of the waiting times $\lambda$ (laminar lengths) within the neighbourhood of the bifurcating fixed point $M=0$ \cite{Contoyiannis2002}. Here, we argue that varying the temperature in the SSB completion region, departing from $T_{pc}$ and approaching $T_{SSB}$, a gradual transition from type I intermittent dynamics to on-off intermittency \cite{Platt1993} takes place. The benchmark of this transition is the collective spin dynamics around the fixed point at $M=0$. For $T \approx T_{pc}$ this dynamics is dominated by the marginal instability of the $M=0$ fixed point leading to a typical type I intermittent behaviour ala Manneville and Pomeau \cite{Manneville1980}. As the temperature approaches $T_{SSB}$ the intermittent behaviour is expressed by long oscillations around the new stable minima interrupted by burst dynamics describing the transition from the neighbourhood of the one stable minimum to the other. The stable minima in $V_{eff}(M)$ represent the stable manifolds in the collective spin dynamics. It is interesting to notice here that similar dynamical behaviour has been observed also in a discrete description of the magnetization dynamics through a non-linear map derived from a Ginzburg-Landau free-energy functional
 \cite{Contoyiannis2021}. The communication between the two minima is achieved through the $M=0$ channel formed by spin configurations in which the clusters with positively aligned spins ($s_i=1$) are equal (in total size) with the clusters with negatively aligned spins ($s_j=-1$). This channel acts as the unstable manifold and becomes more and more narrow (statistically suppressed) as $T \to T_{SSB}$ due to the coalescence of clusters of the same orientation. As a consequence the on-off intermittency becomes established, a behaviour which is expected to be observable in the modification of the laminar length distribution $N(\lambda)$ approaching $T_{SSB}$. As shown in \cite{Cenys1997} for noisy on-off intermittency the laminar length distribution becomes a power-law with exponent $\frac{3}{2}$. We have calculated the laminar length distribution around $M=0$ for the 3d Ising system at the temperatures $T_{pc}=4.515$ and $T=4.45$ which lies very close to $T_{SSB}$ according to the relation in Eq.~(\ref{eq:4}). We pick out in Fig.~4 the bulk of the distribution where the power-law behaviour is clearly seen. The reason for choosing this region of $\lambda$-values is that we want to avoid the saturation region close to $\lambda=1$ (notice that the time is measured in sweeps) as well as the exponential tail of the distribution, induced by the finite size $L$ of the considered system. The increase of the slope of the power-law decay from $\approx 1.2$ to $\approx 1.5$ as $T_{SSB}$ is approached, can be clearly seen.

\begin{figure}[htbp]
\includegraphics[width=0.95\columnwidth]{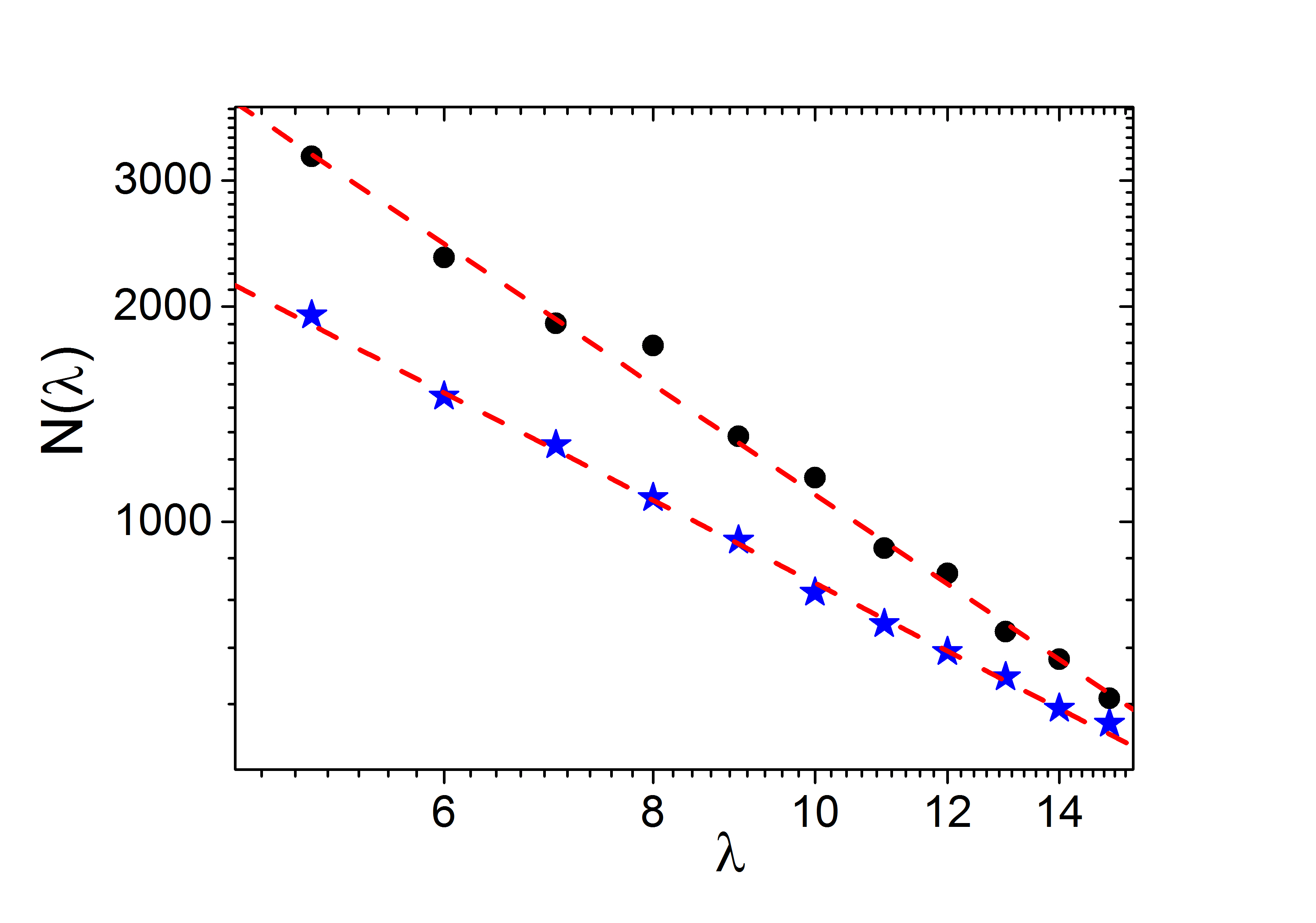}
\caption{The distribution $N(\lambda)$ (in log-log scale) of the waiting times $\lambda$ around the marginally unstable fixed point $M=0$ for the 3d Ising model at $T=4.515$, which is approximately $T_{pc}$ (blue stars) and at $T=4.45$, which is approximately $T_{SSB}$ (black circles). The red dashed lines indicate the power-laws with exponents $1.2$ and $1.58$ (fit results) respectively and they are displayed to guide the eye.}
\label{fig:fig4}
\end{figure}

Related to this transition from type I to on-off intermittency in the magnetization dynamics as the temperature is varied from $T_{pc}$ to $T_{SSB}$ is an anomalous behaviour of the magnetization autocorrelation function displayed in Figs.~5(a,b).

\begin{figure}[htbp]
\includegraphics[width=0.8\columnwidth]{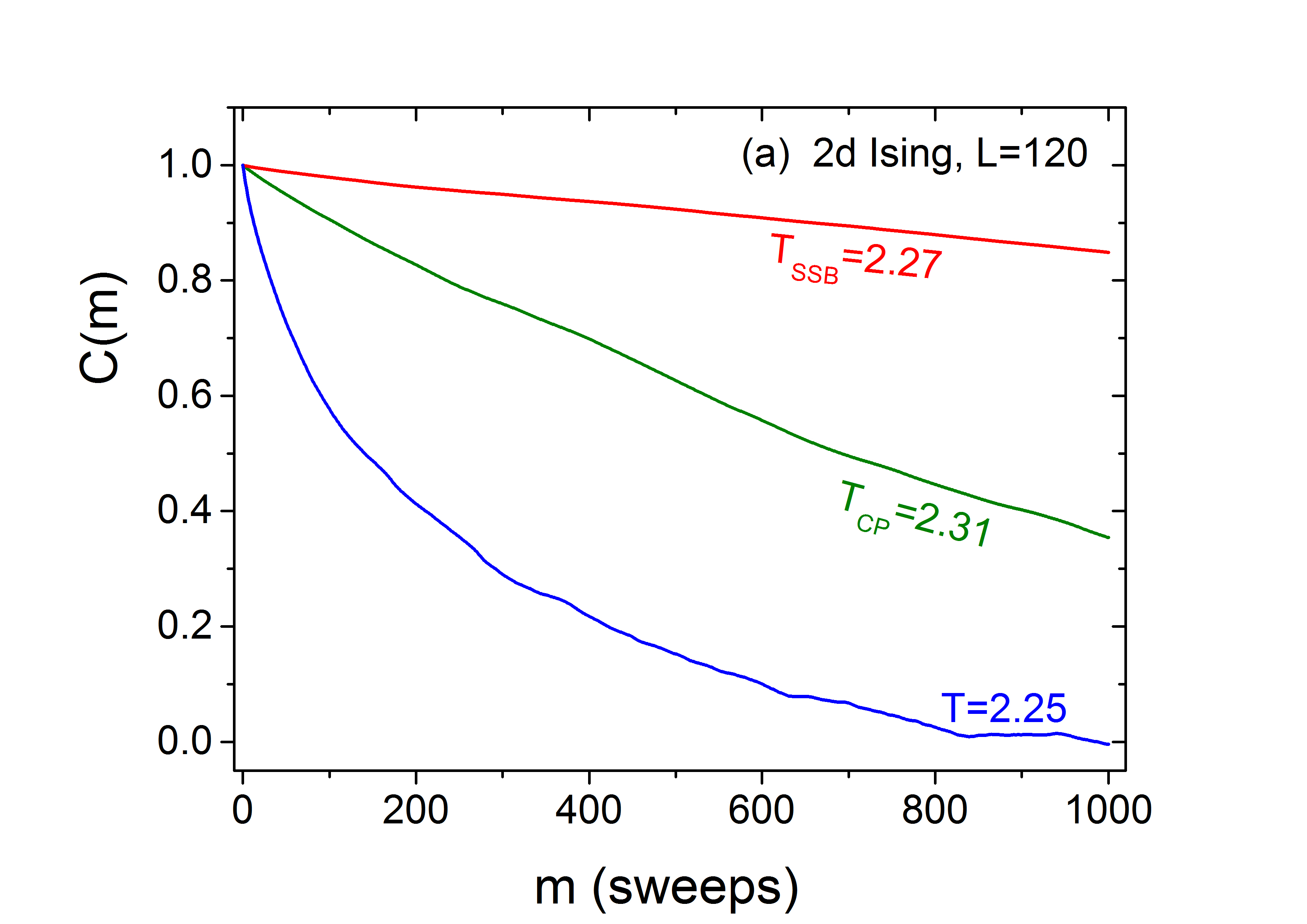}
\includegraphics[width=0.8\columnwidth]{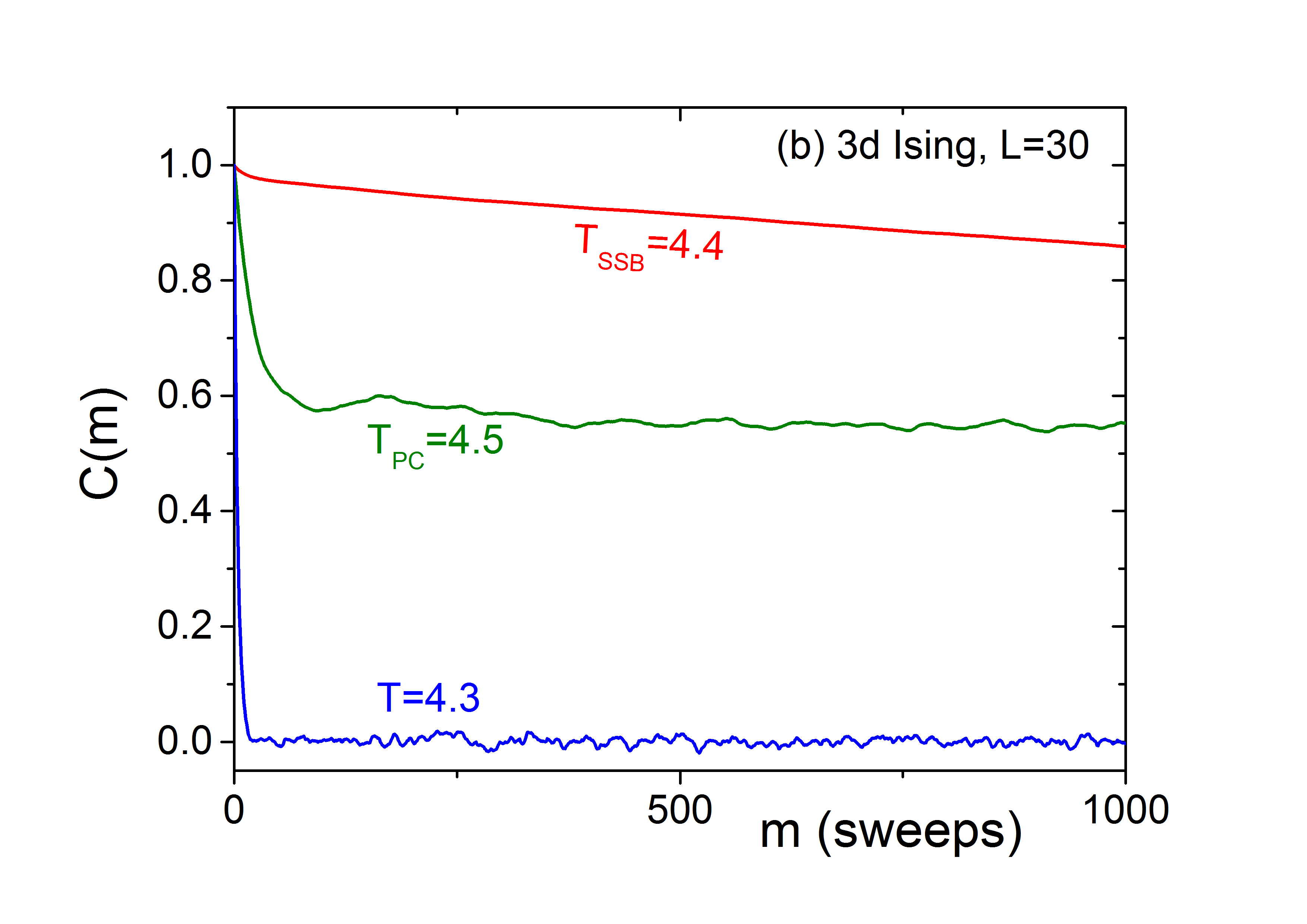}
\caption{In (a) we show the autocorrelation function of the magnetization times-series for the 2d Ising model at three different temperatures $T=2.25$ ($T$ outside ot the SSB completion region), $T_{SSB}=2.27$  and $T_{pc}=2.31$. In (b) it is shown the corresponding quantity for the 3d Ising model now at temperatures $T=4.3$ ($T$ outside ot the SSB completion region), $T_{SSB}=4.4$ and $T_{pc}=4.5$.}
\label{fig:fig5}
\end{figure}

We clearly observe in both cases an increase of the autocorrelations as $T$ approaches $T_{SSB}$. This anomalous enhancement of autocorrelations in the on-off intermittency is due to the fact that, although the mean magnetization in this case remains close to zero, the  magnetization time-series contains mainly small oscillations around the minima of the effective potential $V_{eff}(M)$ (maxima of the distribution $P(M)$) which in turn may lead to large waiting times in this magnetization region.

The analysis presented in this letter can have significant impact on our understanding of critical phenomena as they are realized in true physical systems. In particular, it clarifies the fact that the establishment of a non-vanishing mean value of the order parameter, taking place below $T_{SSB}$, does not coincide with the pseudocritical point $T_{pc}$ of a system where the stable fixed point bifurcates becoming unstable. Only when the, stable in the symmetric phase and unstable in the spontaneously broken phase, fixed point disconnects from the other (stable) fixed points in the order parameter dynamics, the spontaneous breaking of the underlying symmetry is completed. A characteristic property of the SSB completion region is that the distance $T_{pc} - T_{SSB}$ increases with decreasing system's size. As a consequence in a microscopic system this distance can be large. For example, such a case could appear in the Lattice QCD simulations at zero chemical potential. There, the critical point is estimated through the formation of a non-zero $\langle \bar{q} q \rangle$ condensate (sigma condensate), which in fact takes place at $T_{SSB}$. However, since the calculations are performed in quite small lattices (systems) the corresponding pseudocritical temperature could be significantly higher. Such a scenario could explain inconsistencies occurring in the attempt to apply Lattice QCD results to the phenomenology of ultra-relativistic ion collisions, related to the experimental search for the QCD critical endpoint \cite{Gavai2016,Ratti2018}. Even more, assuming that the sigma condensate is formed in the fireball created in the ion collision experiments, it certainly  possesses a finite life time and therefore the arguments developed in the present work should be taken into account when analyzing the recorded data.

\end{document}